\documentclass[preprint,showpacs,preprintnumbers,amsmath,amssymb]
{revtex4}

\usepackage{graphicx}
\usepackage{dcolumn}
\usepackage{bm}

\begin{document}

\preprint{Typeset by REV\TeX~4}

\title{Relativistic heat conduction and thermoelectric
properties of nonuniform plasmas}

\author{M.~Honda}
\affiliation{Plasma Astrophysics Laboratory, Institute for Global Science,
Mie 519-5203, Japan}

\begin{abstract}
Relativistic heat transport in electron-two-temperature plasmas with
density gradients has been investigated.
The Legendre expansion analysis of relativistically modified
kinetic equations shows that strong inhibition of heat flux
appears in relativistic temperature regimes,
suppressing the classical Spitzer-H{\"a}rm conduction.
The Seebeck coefficient, the Wiedemann-Franz law, and
the thermoelectric figure of merit
are derived in the relativistic regimes.
\end{abstract}


\maketitle

The relativistic effects of hot electrons in laboratory plasmas have
attracted much interests in the past few decades \cite{fisch87},
particularly, in the context of the current drive mode in tokamaks and
the electron cyclotron heating in various confinement devices
\cite{bernstein81,rognlien90}.
In the theoretical arena, Braams and Karney (BK) first
presented the relativistic Fokker-Planck equations with the
extended Rosenbluth potentials \cite{braams87},
and applied the equations to derivation of the relativistic
electrical conductivity \cite{braams89}.
Using the BK collision integrals, Shoucri and Shkarofsky
developed a numerical code to survey the relativistic effects on
electron cyclotron wave, fast wave, and lower hybrid current drive mode
in tokamaks \cite{shoucri94}.
Making use of the Chapman-Enskog expansion, Mohanty and Baral
derived relativistic transport coefficients including
magnetic field effects \cite{mohanty96}.

Rapid heating of plasma often leads to a bi-Maxwellian
electron distribution, consisting of a bulk and high-energy tail,
while maintaining relatively cold ions.
In fact, the appearances of two-temperature spectrum
for electrons have been observed in
some experiments with high-intensity lasers \cite{guethlein96}.
The well-pronounced tail and its velocity moment,
which determine the transport properties, are in the
relativistic temperature regime and,
therefore, important to the
fundamental study of relativistic electron transport.

In this Brief Communication, the relativistic transport theory
presented in the previous paper \cite{honda98} is expanded,
aiming at the numerical simulation of high-temperature ignition plasmas,
and celestial plasmas.
We focus on a problem relevant to heat flux
inhibition due to relativistic effects of electrons
within the framework of the
relativistically corrected Spitzer-H{\"a}rm (SH) formula
for electron-two-temperature plasmas with density gradients.
The formula is fully consistent with the
current-neutral condition, so that one can readily couple the
transport coefficients with fluid codes \cite{honda03}.

The relativistic thermal conductivity is derived below,
along the manner developed by BK \cite{braams87,braams89}.
Begin with the Legendre expansion for electron distribution function,
viz., $f({\bf r},{\bf p},t) \simeq f_0({\bf r},p,t)
+({\bf p}/p) \cdot {\bf f}_1({\bf r},p,t)+higher~order~terms$
for the small parameter which is related to a
characteristic field strength \cite{honda98}.
Introducing the relation of $p_x=p {\rm cos}\phi$,
a relativistically extended kinetic equation is averaged over the
solid angle $\Omega$, i.e., $<\cdots>=\int \cdots{\rm d}(\Omega/4\pi)$.
After the manipulations, we obtain the first order equation
in the form of
%\begin{widetext}
\begin{equation}
{\partial f_{1x} \over {\partial t}} + v{\partial f_0 \over {\partial x}}
-{eE_x \over {m_0c}}{\partial f_0 \over {\partial \mu}}
=\left( {\delta f_{1x} \over {\delta t}} \right)_c,
\label{eq:1}
\end{equation}
%\end{widetext}
\noindent
where $v=c \mu/\Gamma$, $\mu=p/m_0c$, and $\Gamma=\sqrt{1+\mu^2}$.
Using the transfer cross section of the relativistic Mott scattering
$\sigma_{ei}^t$, the collision term of
Eq.~(\ref{eq:1}) can be
approximated by $(\delta f_{1x} / {\delta t})_c
\simeq - n_i v \sigma_{ei}^t f_{1x} \equiv -\nu_{ei}f_{1x}$,
where $n_i$ is the number density of ions,
$\nu_{ei}=(n_i Y_{ei}/c^3)(\Gamma/\mu^3)$
is the electron-ion collision frequency,
$Y_{ei}=4 \pi [\bar{Z} e^2/(4 \pi \epsilon_0 m_0)]^2 {\rm ln} \Lambda$,
and $m_0$, $\bar{Z}$, and ${\rm ln} \Lambda$
are the electron rest mass, the averaged charge number,
and the Coulomb logarithm, respectively \cite{honda98}.

In Eq.~(\ref{eq:1}), the effects of magnetic fields are ignored.
This approximation is valid for
$\nu_{ei} \gg \omega_c$, where $\omega_c = |eB|/(\Gamma m_0)$
is the electron cyclotron frequency.
The validity condition gives the allowable parameter range of
the magnetic field strength of
%\begin{widetext}
\begin{equation}
|B| \ll 1.7 \times 10^7
{ \Gamma^2 \over { \left(\Gamma^2 -1 \right)^{3/2} } }
\left( { {\bar{Z}^2 n_i} \over { 10^{27}~\rm{cm}^{-3} } }\right)
\left( { {\rm{ln} \Lambda} \over 10} \right)~\rm{G}.
\label{eq:2}
\end{equation}
%\end{widetext}
\noindent
In highly compressed targets irradiating by a
relativistic laser pulse \citep{tabak94},
the dense plasma parameters are typically
$(\Gamma-1) \sim 10^{-1}$, $n_i \sim 10^{26}~{\rm cm^{-3}}$,
$\bar{Z} \approx 3.5$ (carbonized deuterium-tritium),
and ${\rm ln} \Lambda \approx 5$.
For such parameters, the right-hand side (RHS) of Eq.~(\ref{eq:2}) reads about
$130~{\rm MG}$.
Around the tenuous coronae, the laser pulse drives
relativistic currents, and induces the self-magnetic fields of magnitude
$B \sim m_0 \omega_{pe} c/e \sim 10^2~{\rm MG}$.
Recent numerical simulation indicates that at the surface, an
intense magnetic field of $B \leq 280~{\rm MG}$ prevents hot electrons from
penetrating into the higher density region \citep{sentoku02}.
The electrons near the channel envelope
where the magnetic field is strongest, as well as
lower energy electrons, tend to be magnetically trapped \cite{honda00a},
since their Larmor radii are comparable to or
even less than the channel radius.
This stopping effect seems to be subject
to the Alfv{\'e}n current limit \cite{honda00b}, which
is irrelevant to the limit of energy flux.
For a self-focusing electron beam,
a fraction of energetic electrons, running along the channel axis,
cannot be trapped \cite{honda00a}, and generate the
relatively small magnetic fields of $B<10^2~{\rm MG}$
in the denser plasma, as was shown in Ref.~\citep{sentoku02}.
The penetrating electrons have a largely anisotropic momentum distribution,
which cannot be treated by the diffusion approximation employed here.
In a highly compressed region, however, the transported electrons are
expected to be thermalized via dissipative processes
\cite{honda00b, honda00c},
and the beam type transport may become diffusive,
further decaying the magnetic fields.
Although, diffusive transport plays a significant role in
heating the final compressed fuel,
the details are still not well understood.
Hence, here we investigate the fundamental transport
properties in the parameter regions of the highly compressed ignitor plasma,
where magnetic field effects can be fairly neglected,
as far as Eq.~(\ref{eq:2}) is fulfilled.
I also mention that the density gradient of ablative plasma is likely to be
steep in the higher density regions \cite{honda03},
so that nonuniform effects are taken into account here.

For a quasisteady condition of $\partial f_{1x} / {\partial t} \simeq 0$
in Eq.~(\ref{eq:1}), i.e., omitting the electron inertia,
the anisotropic component of electron distribution function is given by
%\begin{widetext}
\begin{equation}
f_{1x}(x,\mu) \simeq -{c^4 \over {n_i Y_{ei}}}
\left( {\mu^4 \over {1+\mu^2}}{\partial f_0 \over {\partial x}}
-{eE_x \over {m_0 c^2}}{\mu^3 \over {\sqrt{1+\mu^2}}}
{\partial f_0 \over {\partial \mu}} \right).
\label{eq:3}
\end{equation}
%\end{widetext}
\noindent
Heat flux of relativistic electrons can be defined by
$q_x \equiv m_0 c^2 \int_0^{\infty} \int (\Gamma - 1) v_x \mu^2 f
{\rm d}\Omega {\rm d}\mu$.
Integrating over solid angle, yields
$q_x = {4 \over 3} \pi m_0 c^3 \int_0^{\infty} f_{1x} \mu^3
(\Gamma - 1)/\Gamma {\rm d}\mu$.
Making use of Eq.~(\ref{eq:3}), this may be written as
%\begin{widetext}
\begin{equation}
q_x=-{4 \pi m_0 c^7 \over {3 n_i Y_{ei}}}
\left( \int_0^{\infty}
{\mu^7 \over {1+\mu^2}}{\partial f_0 \over {\partial x}}{\rm d}\mu
-{eE_x \over {m_0 c^2}}
\int_0^{\infty} {\mu^6 \over {\sqrt{1+\mu^2}}}
{\partial f_0 \over {\partial \mu}}{\rm d}\mu \right).
\label{eq:4}
\end{equation}
%\end{widetext}
\noindent
The longitudinal electric field $E_x$ in Eq.~(\ref{eq:4}) can be
determined by the current-neutral condition
$j_x \equiv -e \int_0^{\infty} \int v_x \mu^2 f 
{\rm d}\Omega {\rm d}\mu =
-{4 \over 3} \pi ec \int_0^{\infty} f_{1x} \mu^3/\Gamma {\rm d}\mu \simeq 0$.
That is,
%\begin{widetext}
\begin{equation}
{eE_x \over {m_0 c^2}}=
\frac {{
 \int_0^{\infty}
{\frac {\mu^7}{(1+\mu^2)^{3/2}}}
{\frac {\partial f_0}{\partial x}}}{\rm d}\mu
}{
\int_0^{\infty}
{\frac {\mu^6}{1+\mu^2}}
{\frac {\partial f_0}{\partial \mu}}{\rm d}\mu
}.
\label{eq:5}
\end{equation}
%\end{widetext}
\noindent
For the isotropic component $f_0(x,\mu)$,
I employ the superposition of the two-temperature populations of electrons,
%\begin{widetext}
\begin{equation}
f_0(x,\mu) = {1\over {4\pi}} \sum_{j}
{ n_{e,j}(x) \alpha_j(x) \over {K_2[\alpha_j(x)]}}
{\rm exp}\left[-\alpha_j(x) \sqrt{1+\mu^2}\right],
\label{eq:6}
\end{equation}
%\end{widetext}
\noindent
where $K_{\nu}(\alpha_j)$ is the modified Bessel function of index $\nu$
with its argument of $\alpha_j(x) \equiv m_0c^2/T_j(x)$, and
$j=c, h$ indicate the cold and hot components, respectively.
The normalization is given by
$n_e(x)=\bar{Z} n_i(x)=\sum_{j} n_{e,j}(x)=4\pi \int_0^{\infty}
f_0(x,\mu) \mu^2 {\rm d}\mu$.

In a steep temperature gradient plasma, depending on the collisional
mean-free path, $\lambda$,
the transport properties may not be locally defined.
In this sense, local transport theory is valid only for
the case of $\lambda \ll |L_T|$, where $L_T=T/(\partial T/\partial x)$ is the
characteristic length of the temperature gradient.
Concerning the relation of $|\partial T/\partial x| \sim e|E_x|$
derived from Eq.~(\ref{eq:5}),
the parameter range involving the electric field can be estimated as
$
|E_x| \ll E_c \sim 10^{12}
( {100~\rm{keV}} / T )
( {\bar{Z}^2 n_i} / { 10^{27}~\rm{cm}^{-3} } )
( {\rm{ln} \Lambda} / 10 )~\rm{V/m}.
$
In the case of ignitor physics,
the relativistic electron transport establishes the temperature gradient
in the high-density plasma.
Assuming the spatial gradient of
$\Delta T /\Delta x \sim -100~{\rm keV}/100~{\rm \mu m}$,
the electric field strength can be estimated as
$|E_x| \sim 10^{9}~{\rm V/m}$ [see also Eq.~(\ref{eq:12}) below].
For $T \leq 10^2~{\rm keV}$, $\bar{Z}^2 n_i \sim 10^{27}~{\rm cm^{-3}}$,
and ${\rm ln} \Lambda \approx 5$, we read $|E_x| < 10^{-2} E_c$.
For the case of $|E_x| > (0.01-0.1)E_c$,
one may solve kinetic transport equations
to determine the full self-consistent spectral distribution,
instead of using Eq.~(\ref{eq:6}) \cite{bell81}.

Substituting Eq.~(\ref{eq:6}) into Eqs.~(\ref{eq:4}) and (\ref{eq:5}),
we obtain the relativistic heat flux of
$q_x \equiv -\kappa_{rel}(\partial T_h/\partial x)$
for the temperature gradient of hot electrons,
and may decompose the coefficient as
$\kappa_{rel}=f_{rel}\kappa_{\rm SH}$.
Here, $\kappa_{\rm SH}(T_h) = 256(2 \pi)^{1/2} \epsilon_0^2 T_h^{5/2}
/(\bar{Z} e^4 m_0^{1/2} {\rm ln}{\rm \Lambda})$
is the familiar nonrelativistic SH heat conductivity of the Lorentz plasmas
\cite{spitzer53}, and the factor $f_{ref}$ corresponds to the
relativistically corrected flux limiter which can be expressed as
%\begin{widetext}
\begin{equation}
f_{rel}=
{(2\pi)^{1/2} \over 384} \alpha_h^{7/2}
\{
C_{1,c}+\theta C_{1,h}+\epsilon
\left[
C_{2,c}\Theta_1(\alpha_c)+C_{2,h}\Theta_1(\alpha_h)
\right]
\},
\label{eq:8}
\end{equation}
%\end{widetext}
\noindent
where the abbreviations are
\begin{equation}
\epsilon=
\frac{
\theta C_{2,c}
\left[\Theta_1(\alpha_c)+C_{3,c}\Theta_2(\alpha_c) \right]
+C_{2,h}
\left[\Theta_1(\alpha_h)+C_{3,h}\Theta_2(\alpha_h) \right]
}{
{\alpha_c}C_{2,c} \Theta_2 (\alpha_c)
+{\alpha_h}C_{2,h} \Theta_2 (\alpha_h)
};\label{eq:9}
\end{equation}
\begin{subequations}
\label{eq:10}
\begin{equation}
C_{1,j}=-\frac{
C_{2,j} \left[ C_{3,j} \Theta_1(\alpha_j) - \Theta_3(\alpha_j) \right]
}{\alpha_j},\label{subeq:a}
\end{equation}
\begin{equation}
C_{2,j}={\frac {n_{e,j}}{n_e}}
{\frac {\alpha_j}{K_2(\alpha_j)}},\label{subeq:b}
\end{equation}
\begin{equation}
C_{3,j}=3 - \delta_j + \frac {\alpha_j K_1(\alpha_j)}{K_2(\alpha_j)}
;\label{subeq:c}
\end{equation}
\end{subequations}
\begin{subequations}
\label{eq:11}
\begin{equation}
\Theta_1(\alpha_j)=\left(
1 - {1\over \alpha_j} + {2\over \alpha_j^2}
+ {42\over \alpha_j^3} + {120\over \alpha_j^4}
+ {120\over \alpha_j^5} \right) {\rm exp}(-\alpha_j)
+ \alpha_j {\rm Ei}(-\alpha_j),\label{subeq:a}
\end{equation}
\begin{equation}
\Theta_2(\alpha_j)=\left(
1 - {1\over \alpha_j} + {2\over \alpha_j^2}
- {6\over \alpha_j^3} - {24\over \alpha_j^4} - {24\over \alpha_j^5}
\right) {\rm exp}(-\alpha_j)
+ \alpha_j {\rm Ei}(-\alpha_j),\label{subeq:b}
\end{equation}
\begin{equation}
\Theta_3(\alpha_j)=\left(
{48\over \alpha_j^2} + {288\over \alpha_j^3}
+ {720\over \alpha_j^4} + {720\over \alpha_j^5} \right)
{\rm exp}(-\alpha_j),\label{subeq:c}
\end{equation}
\end{subequations}
\noindent
where ${\rm Ei}(-\alpha_j)$ is the exponential integral function,
and $\delta_j = \partial {\rm ln} n_{e,j}/\partial {\rm ln} T_j$
and $\theta = \partial {\rm ln} T_c/\partial {\rm ln} T_h$
reflect the nonuniformity of plasma.
Namely, for $\theta \rightarrow 0$, the formula describes the
energetic transport in the plasma that
the cold electron component is isothermal,
and for $\delta_j \rightarrow 0$ and $-1$, 
in the plasma that the electron component $j$ is
isochoric ($n_{e,j}={\rm const}$) and
isobaric ($n_{e,j}T_j={\rm const}$), respectively.

The geometrical constraint of $\nabla n_{e,j} \parallel \nabla T_j$
due to ignoring two-dimensional (2D) effects
means that thermoelectric magnetic fields,
which can be prominent,
for example, in intense laser-plasma interactions \cite{borghesi98},
are not taken into account at the moment.
2D effects are important, because
they prefer to short out electric fields and pinch directional flows by
the toroidal magnetic fields.
The complexities of magnetic inhibition in heat flux
might be effectively considered by introducing a reduction factor
$f_{\rm B} < 1$: Bohm's $f_{\rm B} = (1 + \omega_c \tau )^{-1}$
or Braginskii's $f_{\rm B} = (1 + \omega_c^2 \tau^2 )^{-1}$
\cite{borghesi98}, where $\tau$ denotes a collision period.
That is, one can practically utilize the cross-field conductivity
approximated by $\kappa_{\perp rel} \approx f_{\rm B} f_{rel} \kappa_{\rm SH}$.
Note that Eq.~(\ref{eq:2}) reflects the much smaller Hall parameter
$\omega_c \tau \ll 1$, such that $f_{\rm B} \rightarrow 1$, and
$\kappa_{\perp rel} \simeq \kappa_{\parallel rel} = \kappa_{rel}$.

In the following, more elemental issues are investigated, i.e.,
relativistically extended longitudinal thermoelectric effects.
With regard to the longitudinal thermal diffusion
that develops an electrostatic potential,
one should note the important relation
$\epsilon = L_{T_h}[eE_x/(m_0c^2)]$.
For the special case of $n_{e,h}/n_e \rightarrow 1$,
$C_{2,c} \rightarrow 0$, and $\theta \rightarrow 1$,
namely, the one-temperature model for electrons,
the self-consistent electric field Eq.~(\ref{eq:9}) reduces to
%\begin{widetext}
\begin{equation}
\epsilon \simeq {\frac {1}{\alpha}}
\left[ \frac {\Theta_1(\alpha)}{\Theta_2(\alpha)}
+ C_3(\alpha,\delta)
\right],\label{eq:12}
\end{equation}
%\end{widetext}
\noindent
where $\alpha \equiv \alpha_j$ and $C_3 \equiv C_{3,j}$.
In the thermoelectric point of view,
the relativistic Seebeck coefficient can be
defined by $s \equiv \alpha \epsilon/e$.
The temperature dependence of Eq.~(\ref{eq:12}) is shown in
Table~\ref{tab:table1} for $\delta \equiv \delta_j = 0$ and $-1$.
In the nonrelativistic limit of $\alpha \gg 1$,
Eq.~(\ref{eq:12}) for $\delta = 0$ asymptotically approaches
$\epsilon \rightarrow -5/(2 \alpha)$ \cite{honda98}.
In the isobaric case of $\delta = -1$,
owing to the pressure-balance effects, the field strength reduces to
$50-60~\%$ of the $\delta=0$ case.
Noted is that in this case the flux limiter $f_{rel}$
does not depend on $\delta$, and the similar property appears
again in the following other cases.

\begin{figure*}

\resizebox{150mm}{!}{\includegraphics{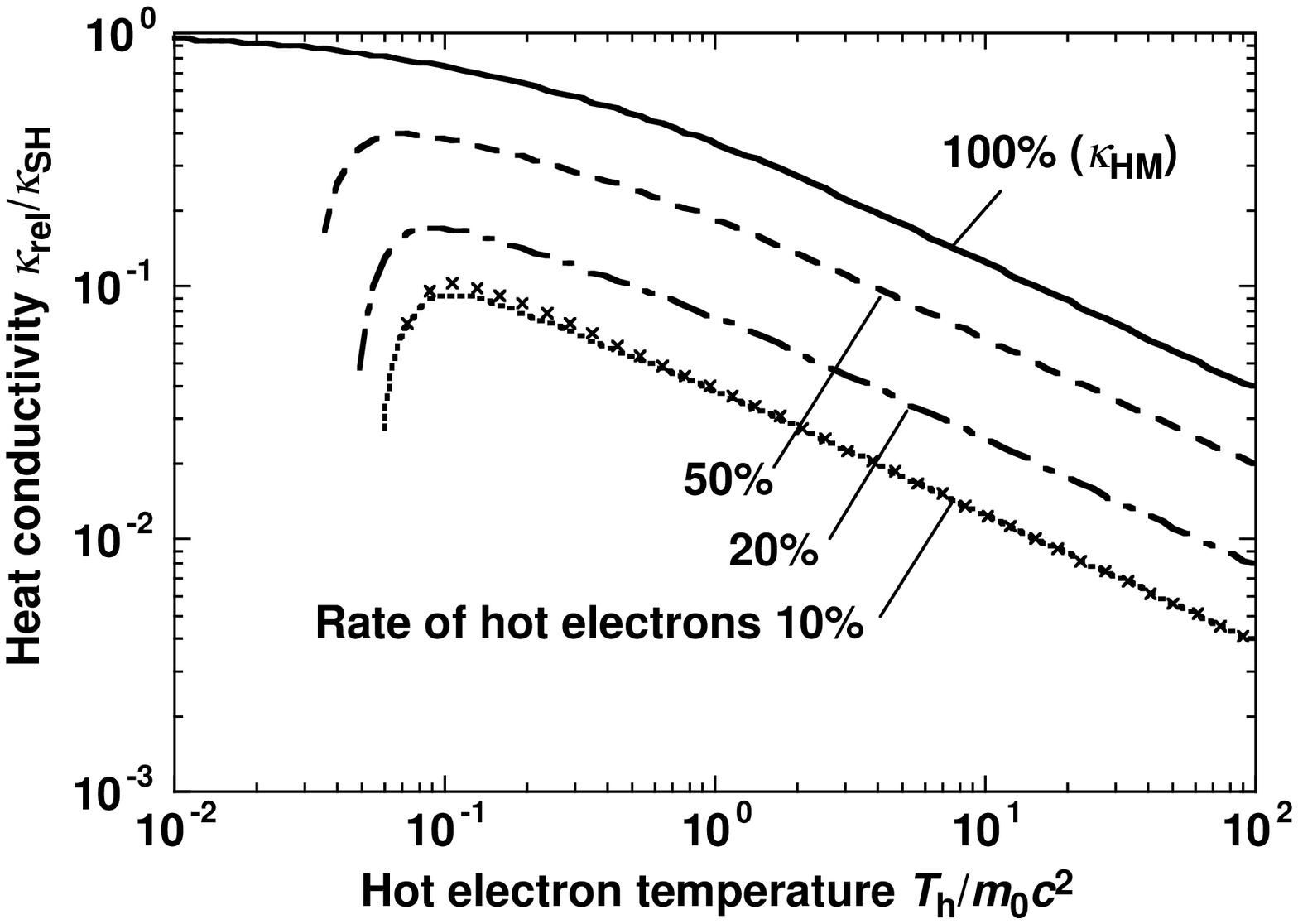}}
\caption{\label{fig:fig1}
Hot electron temperature dependence of the relativistic
Spitzer-H{\"a}rm (SH) heat conductivity normalized by the
nonrelativistic one:
The flux limiter of hot electrons
$f_{rel}(T_h)=\kappa_{rel}/\kappa_{\rm SH}$ is shown for
density ratios of $n_{e,h}/n_e = 1$
[Eq.~(\ref{eq:13}): solid curve], $0.5$ (dashed curve),
$0.2$ (dotted-dashed curve), and $0.1$ (dotted curve).
For the case of $n_{e,h}/n_e \neq 1$,
$T_c/m_0c^2=0.01$ and $n_{e,j}={\rm const}$ are chosen as examples.
For comparison, we plot another case of
$n_{e,j} \propto T_j^{-1}$, only for $n_{e,h}/n_e=0.1$ and
$T_c/m_0c^2 = 0.01$ (crosses).
}
\end{figure*}

In Fig.~\ref{fig:fig1}, the temperature dependence of the relativistically
corrected flux limiter are shown.
The ratios of hot/total electron density are chosen for $n_{e,h}/n_e=0.1-1$,
fixing the temperature scale length equal, $\theta=1$.
For the electron-two-temperature models of
$n_{e,h}/n_e \neq 1$ ($C_{2,c} \neq 0$), set the temperature
of cold component to $\alpha_c=10^2$ ($T_c = 5.11~{\rm keV}$) as an example.
The densities are set to be uniform ($\delta_j=0$),
except for the case of $n_{e,h}/n_e=0.1$ that
the nonuniformity ($\delta_j=-1$) is taken into consideration.
Actually our major interests are in the relativistic heat flux
carried by the high-energy tail electrons of
$\Gamma > \Gamma_0 = (\alpha_h - \alpha_c)^{-1}
{\rm ln}(C_{2,c}/C_{2,h})$, where the spectral population of
hot electrons is larger than that of cold ones.
As expected, in the lower energy regions of $\Gamma < \Gamma_0$,
energy transport by cold components is dominant.
In Fig.~\ref{fig:fig1}, such criterion seems to appear as
pseudo cut-off in the lower temperature region.

Now one finds that the heat flux is strongly inhibited
in the relativistic regime.
For example, in the one-temperature model for electrons,
the flux limiters are $f_{rel} \simeq 0.73$
for $T=0.1 m_0 c^2$ ($\alpha=10$) and
$f_{rel} \simeq 0.37$ for $T=m_0 c^2$ ($\alpha=1$),
as shown in Fig.~\ref{fig:fig1} (solid curve).
This is due to the drift velocity
carrying heat asymptotically close to the speed of light.
Moreover, it is found that a fall in the hot electron population
leads to further decrease of the conductivity, and indeed,
the degree of the depletion reflects the abundance of hot electrons.
Regarding the electron transport in laser-produced plasmas,
typically a flux limiter of order of $10^{-2}-10^{-1}$
has been empirically employed \cite{bell81},
consistent with the experimental results \cite{malone75}.
In this aspect, the present results imply that the
relativistic effects on sparsely populated high-energy tails
can also participate in lowering the flux limit.
These properties do not largely depend on $\delta_j$
as seen in Fig.~\ref{fig:fig1}.
For example, in the case of $n_{e,h}/n_e = 0.1$,
the difference of the flux limiter between the case of
$\delta_j = 0$ (dotted curve) and $-1$ (crosses) is about
$10\%$ at most.

Taking the limit of $n_{h,e}/n_e \rightarrow 1$,
$C_{2,c} \rightarrow 0$, and $\theta \rightarrow 1$,
Eqs.~(\ref{eq:8})-(\ref{eq:11}) reduce to
%\begin{widetext}
\begin{equation}
f_{rel} \simeq
{\frac {(2\pi)^{1/2}}{384}}
{\frac {\alpha^{7/2}}{K_2(\alpha)}}
\left[ \frac {\Theta_1^2(\alpha)}{\Theta_2(\alpha)}
+ \Theta_3(\alpha)
\right].\label{eq:13}
\end{equation}
%\end{widetext}
\noindent
This corresponds to the standard relativistic SH heat conductivity
having the temperature dependence of
$\kappa_{\rm HM}(T)=f_{rel}(T) \kappa_{\rm SH}(T) \propto T^2-T^{5/2}$
\cite{honda98},
which exhibits the asymptotic property of
$\kappa_{\rm SH}(T) \propto T^{5/2}$ ($f_{rel}\rightarrow 1$)
in the nonrelativistic limit of $\alpha \gg 1$ \cite{spitzer53},
whereas $\kappa_{\rm DT}(T)=[5 (2\pi)^{1/2}/32] \alpha^{1/2}
\kappa_{\rm SH}(T) \propto T^2$ ($f_{rel}\propto T^{-1/2}$)
by Dzhavakhishvili and Tsintsadze in the ultrarelativistic limit of
$\alpha \ll 1$ \cite{dzhavakhishvili73}.
These characteristics are also shown in Fig.~\ref{fig:fig1}
(solid curve), and summarized in Table~\ref{tab:table1}.

\begin{figure*}
\resizebox{140mm}{!}{\includegraphics{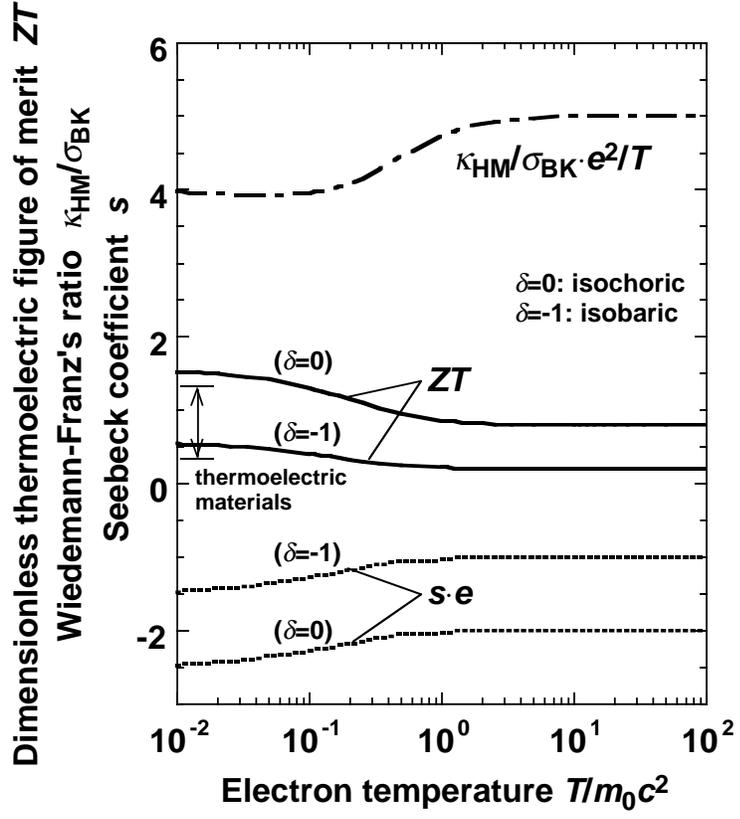}}
\caption{\label{fig:fig2}
Electron temperature dependence of the dimensionless
thermoelectric figure of merit $ZT$ (solid curves),
the Wiedemann-Franz ratio $\kappa_{\rm HM}/\sigma_{\rm BK}$
multiplied by $e^2/T$ (dotted-dashed curve), and the
Seebeck coefficient $s$ multiplied by $|e|$ (dotted curves).
Note that in the present case
$\kappa_{\rm HM}/\sigma_{\rm BK}$ is independent on
$\delta$ and takes the minimum values at
$T/m_0c^2 \simeq 0.051$.
}
\end{figure*}

Here let us take the ratio of the thermal to electrical conductivity.
The key relation is known as the Wiedemann-Franz law
for metallic states of matters \cite{landau81}.
The ubiquitous nature is derived from a simple assumption of the
elastic scattering of conduction electrons.
As for fully ionized plasmas, the relativistically extended law
can be expressed as
%\begin{widetext}
\begin{equation}
{\frac {\kappa_{\rm HM}}{\sigma_{\rm BK}}} =
-{\frac {T}{e^2}}
{\frac {\Theta_1^2(\alpha)+\Theta_2(\alpha)\Theta_3(\alpha)}
{\Theta_2^2(\alpha)}}>0
,\label{eq:14}
\end{equation}
%\end{widetext}
\noindent
for the case of $n_{e,h}/n_e = 1$.
Here, $\sigma_{\rm BK}(\alpha)=-[(2\pi)^{1/2}/96]
[\alpha^{7/2} \Theta_2(\alpha)/K_2(\alpha)]
\sigma_{\rm S} > 0$ \cite{braams89},
and $\sigma_{\rm S}(T)=64(2\pi)^{1/2} \epsilon_0^2 T^{3/2}/
(\bar{Z} e^4 m_0^{1/2} {\rm ln \Lambda})$
stands for the nonrelativistic Spitzer conductivity.
Evidently, the ratio depends on the temperature only,
without involving the intrinsic parameters of plasmas.
Figure~\ref{fig:fig2} shows the temperature dependence of Eq.~(\ref{eq:14}).
In the nonrelativistic limit of $\alpha \gg 1$,
it asymptotically approaches
$\kappa_{\rm HM}/\sigma_{\rm BK} \rightarrow
\kappa_{\rm SH}/\sigma_{\rm S} = 4T/e^2$ \cite{spitzer53}.
This value slightly decreases as the temperature increases,
to take the minimum value of
$(\kappa_{\rm HM}/\sigma_{\rm BK})_{\rm min} = 3.92T/e^2$
at $\alpha = 19.6$ ($T = 26.1~{\rm keV}$).
As seen in the figure, it increases up to
$\kappa_{\rm DT}/\sigma_{\rm BK} = 5T/e^2$ in the ultrarelativistic regime.
It may be instructive to mention that the transport equation of the
Fermi liquid in metals or condensed plasmas yields
$\kappa/\sigma \simeq \pi^2 T/(3e^2) \simeq 3.3T/e^2$
\cite{landau81,lee84}, which is lower than
$(\kappa_{\rm HM}/\sigma_{\rm BK})_{\rm min}$ in the ordinary plasmas.

The heat conductivity holds the larger power index of temperature.
Thus, fast heating of plasma can drive the nonlinear heat-wave
accompanied with a well-defined wave front, where an
electrostatic field tends to be well developed.
This leads to an idea that such a thermally non-equilibrated plasma
can be essentially compared to a thermoelectric converter.
And, in general, its efficiency can be quantitatively evaluated
by invoking a thermoelectric figure of merit.
Along the conventional notation used in material physics,
we now define the thermoelectric figure of merit by
$Z \equiv s^2 \sigma_{\rm BK}/\kappa_{\rm HM}=(\alpha \epsilon/e)^2
(\sigma_{\rm BK}/\kappa_{\rm HM})$ for $n_{e,h}/n_e = 1$.
Making use of Eqs.~(\ref{eq:12}) and (\ref{eq:14}),
this multiplied by $T$ can be written in the dimensionless form,
%\begin{widetext}
\begin{equation}
ZT=-\frac {
\left[ \Theta_1(\alpha) + \Theta_2(\alpha)C_3(\alpha,\delta)
\right]^2
}
{\Theta_1^2(\alpha)+\Theta_2(\alpha)\Theta_3(\alpha)} > 0
.\label{eq:15}
\end{equation}
%\end{widetext}
\noindent
Note that Eq.~(\ref{eq:15}) depends on $\delta$,
in contrast to Eqs.~(\ref{eq:13}) and (\ref{eq:14}).
For $\delta=0$ and $-1$, the temperature dependence of Eq.~(\ref{eq:15})
and the coefficient $s$ are also shown in Fig.~\ref{fig:fig2}.
It is found that for $\alpha \gg 1$, Eq.~(\ref{eq:15})
asymptotically approaches $ZT(\delta = 0) \rightarrow {25 \over 16}$
and $ZT(\delta = -1) \rightarrow {9 \over 16}$,
whereas for $\alpha \ll 1$, $ZT(\delta = 0) \rightarrow {4 \over 5}$
and $ZT(\delta = -1) \rightarrow {1 \over 5}$.
Particularly, in the nonrelativistic plasmas with uniform density,
i.e., $\alpha \gg 1$ and $\delta=0$,
one can extract the higher figure of merit $ZT = 1.56$,
compared with the typical thermoelectric materials which provide
$ZT \simeq 0.4-1.3$ as indicated by the arrow in Fig.~\ref{fig:fig2}
\cite{slack95}.
Notice that the Carnot efficiency can be achieved for $ZT \gg 1$.
The dimensionless values of
$(\kappa_{\rm HM}/\sigma_{\rm BK})(e^2/T)$, $(se)^2$, and $ZT$
for some $\alpha$ and $\delta$ values are summarized in Table~\ref{tab:table2}.

In conclusion, I have derived solutions for the
heat conductivity and related thermoelectric coefficients
in a relativistic nonuniform plasma.
These results indicate that the relativistic effects on the
high-energy tail electrons significantly limit the heat flux.
This mechanism might play an additional role of the stopping
of relativistic electrons in the context of
ignitor physics \cite{honda00c},
although this work ignores 2D thermoelectric effects such as
$\nabla_{\perp} n \times \nabla_{\parallel} T$,
which may be important for typical ignitor geometries.

\clearpage

\noindent\\

\clearpage
\begin{table*}
\caption{\label{tab:table1}
Temperature dependence of $f_{rel}$ and $\alpha \epsilon$ for
$\delta=0$ and $-1$.}
\begin{ruledtabular}
\begin{tabular}{rclll}
 $\alpha$ & $T$~(${\rm keV}$)~ & ${\alpha \epsilon}|_{\delta = 0}$ &
 ${\alpha \epsilon}|_{\delta = -1}$ & ~~~$f_{rel}$\footnotemark[1]\\ \hline
 $\ll 1$ & & $-2$ & $-1$ & \\
 $1$ & $5.11 \times 10^2$ & $-2.0178$ & $-1.0178$ & $0.36792$\\
 $5$ & $1.02 \times 10^2$ & $-2.1692$ & $-1.1692$ & $0.62668$\\
 $10$ & $51.1$ & $-2.2692$ & $-1.2692$ & $0.73598$\\
 $20$ & $25.5$ & $-2.3560$ & $-1.3560$ & $0.83044$\\
 $100$ & $5.11$ & $-2.4638$ & $-1.4638$ & $0.95529$\\
 $\gg 1$ & & $-2.5$ & $-1.5$ & $1$\\
\end{tabular}
\end{ruledtabular}
\footnotetext[1]{$f_{rel}=\kappa_{\rm HM}/\kappa_{\rm SH}$ \cite{honda98},
to give $f_{rel} \simeq 1$ \cite{spitzer53} and $\propto T^{-0.5}$
\cite{dzhavakhishvili73} for $\alpha \gg 1$ and $\ll 1$, respectively.}
\end{table*}

\clearpage
\begin{table*}
\caption{\label{tab:table2}
Temperature dependence of $(\kappa_{\rm HM}/\sigma_{\rm BK})(e^2/T)$,
$(se)^2$, and $ZT$ for $\delta=0$ and $-1$.}
\begin{ruledtabular}
\begin{tabular}{cclllll}
 $\alpha$~ &
 $T$~(${\rm keV}$) &
 $(\kappa_{\rm HM}/\sigma_{\rm BK})(e^2/T)$\footnotemark[1] &
 $(se)^2|_{\delta = 0}$ & $(se)^2|_{\delta = -1}$ &
 $ZT|_{\delta = 0}$ & $ZT|_{\delta = -1}$\\ \hline
 $\ll 1$ & $-$ & ~~~~~~~$5$ & $4$ & $1$ & $4/5$ & $1/5$\\
 ~~$0.5$ & $1.02 \times 10^3$ & ~~~~~~~$4.9032$ & $4.0145$ & $1.0072$
 & $0.81874$ & $0.20542$\\
 ~~$1$~~ & $5.11 \times 10^2$ & ~~~~~~~$4.7331$ & $4.0717$ & $1.0360$
 & $0.86026$ & $0.21888$\\
 ~~$5$~~ & $1.02 \times 10^2$ & ~~~~~~~$4.0886$ & $4.7053$ & $1.3670$
 & $1.1508$ & $0.33434$\\
 ~$10$~~ & $51.1$ & ~~~~~~~$3.9529$ & $5.1495$ & $1.6110$ & $1.3027$
 & $0.40754$\\
 ~$20$~~ & $25.5$ & ~~~~~~~$3.9221$ & $5.5509$ & $1.8388$ & $1.4153$
 & $0.46883$\\
 $100$~~~ & $5.11$ & ~~~~~~~$3.9669$ & $6.0704$ & $2.1428$ & $1.5303$
 & $0.54170$\\
 $\gg 1$ & $-$ & ~~~~~~~$4$ & $25/4$ & $9/4$ & $25/16$ & $9/16$\\
\end{tabular}
\end{ruledtabular}
\footnotetext[1]{$\sigma_{\rm BK}(T)$ is introduced in Ref.~\cite{braams89}.}
\end{table*}


\clearpage
\begin{thebibliography}{99}

\bibitem{fisch87}
N.~Fisch, Rev.~Mod.~Phys. {\bf 59}, 175 (1987).

\bibitem{bernstein81}
I.~Bernstein and D.~C.~Baxter, Phys.~Fluids {\bf 24}, 108 (1981).

\bibitem{rognlien90}
T.~D.~Rognlien, Y.~Matsuda, B.~W.~Stellard, and J.~J.~Stewart,
Phys.~Fluids B {\bf 2}, 338 (1990).

\bibitem{braams87}
B.~J.~Braams and C.~F.~F. Karney, Phys.~Rev.~Lett. {\bf 59}, 1817 (1987).

\bibitem{braams89}
B.~J.~Braams and C.~F.~F. Karney, Phys.~Fluids B {\bf 1}, 1355 (1989).

\bibitem{shoucri94}
M.~Shoucri and I.~Shkarofsky, Comput.~Phys.~Comm. {\bf 82}, 287 (1994).

\bibitem{mohanty96}
J.~N.~Mohanty and K.~C.~Baral, Phys.~Plasmas {\bf 3}, 804 (1996).

\bibitem{guethlein96}
G.~Guethlein, M.~E.~Foord, and D.~Price, Phys.~Rev.~Lett. {\bf 77},
1055 (1996).

\bibitem{honda98}
M.~Honda and K.~Mima, J.~Phys.~Soc.~Jpn. {\bf 67}, 3420 (1998).

\bibitem{honda03}
M.~Honda, Jpn.~J.~Appl.~Phys. {\bf 42}, 5280 (2003).

\bibitem{tabak94}
M.~Tabak, J.~Hammer, M.~E.~Glinsky, W.~L.~Kruer, S.~C.~Wilks,
J.~Woodworth, E.~M.~Campbell, and M.~D.~Perry,
Phys.~Plasmas {\bf 1}, 1626 (1994).

\bibitem{sentoku02}
Y.~Sentoku, K.~Mima, Z.~M.~Sheng, P.~Kaw, K.~Nishihara, and
K.~Nishikawa, Phys.~Rev.~E, {\bf 65}, 046408 (2002).

\bibitem{honda00a}
M.~Honda, Phys.~Plasmas, {\bf 7}, 1606 (2000).

\bibitem{honda00b}
M.~Honda, J.~Meyer-ter-Vehn, and A.~Pukhov,
Phys.~Plasmas, {\bf 7}, 1302 (2000).

\bibitem{honda00c}
M.~Honda, J.~Meyer-ter-Vehn, and A.~Pukhov,
Phys.~Rev.~Lett. {\bf 85}, 2128 (2000).

\bibitem{bell81}
A.~R.~Bell, R.~G.~Evans, and D.~J.~Nicholas,
Phys.~Rev.~Lett. {\bf 46}, 243 (1981);
G.~J.~Rickard, A.~R.~Bell, and E.~M.~Epperlein,
{\it ibid.} {\bf 62}, 2687 (1989).

\bibitem{spitzer53}
L.~Spitzer and R.~H{\"a}rm, Phys.~Rev. {\bf 89}, 997 (1953).

\bibitem{borghesi98}
M.~Borghesi, A.~J.~MacKinnon, A.~R.~Bell, R.~Gaillard, and O.~Willi,
Phys.~Rev.~Lett. {\bf 81}, 112 (1998).

\bibitem{malone75}
R.~C.~Malone, R.~L.~McCrory, and R.~L.~Morse,
Phys.~Rev.~Lett. {\bf 34}, 721 (1975).

\bibitem{dzhavakhishvili73}
D.~I.~Dzhavakhishvili and N.~L.~Tsintsadze, Sov.~Phys.~JETP {\bf 37},
666 (1973).

\bibitem{landau81}
L.~D.~Landau and E.~M.~Lifshitz, {\it Physical Kinetics}
(Pergamon,~Oxford,~1981).

\bibitem{lee84}
Y.~T.~Lee and R.~M.~More, Phys.~Fluids {\bf 27}, 1273 (1984).

\bibitem{slack95}
G.~A.~Slack, in {\it CRC Handbook of Thermoelectrics},
edited by D.~M.~Rowe (Chemical Rubber Company,~Boca Raton, FL, 1995).

\end{thebibliography}
\end{document}